\documentclass[12pt,journal,onecolumn,draftcls,peerreview]{IEEEtran}
\usepackage{amssymb}
\usepackage{amsmath}
\usepackage{amsfonts}
\usepackage{graphicx}
\usepackage{algorithm}
\usepackage{algorithmic}
\usepackage{cite}
\usepackage{epstopdf}
\usepackage{multirow}
\usepackage{stfloats}
\usepackage{url}

\begin{document}

\title{The three primary colors of mobile systems}
\author{Hui Liu, ~\IEEEmembership{Fellow,~IEEE,} Zhiyong Chen, ~\IEEEmembership{Member,~IEEE,} Liang Qian
\thanks{The authors are with the Institute of Wireless Communications Technology (IWCT), Department of Electronic Engineering, Shanghai Jiao Tong University, Shanghai, P. R. China, also with Cooperative Medianet Innovation Center. Emails: {\{huiliu, zhiyongchen, lqian\}@sjtu.edu.cn.} Z. Chen and Q. Liang as both the corresponding authors.}}
\maketitle

\begin{abstract}
In this paper, we present the notion of \textquotedblleft mobile 3C systems" in which the \textquotedblleft Communications", \textquotedblleft Computing", and \textquotedblleft Caching" (i.e., 3C) make up the three \textbf{\emph{primary}} resources/funcationalties, akin to the three primary colors, for a mobile system. We argue that in future mobile networks, the roles of computing and caching are as intrinsic and essential as communications, and only the collective usage of these three primary resources can support the sustainable growth of mobile systems. By defining the 3C resources in their canonical forms, we reveal the important fact that \textquotedblleft caching" affects the mobile system performance by introducing non-causality into the system, whereas \textquotedblleft computing" achieves capacity gains by performing logical operations across mobile system entities. Many existing capacity-enhancing techniques such as coded multicast, collaborative transmissions, and proactive content pushing can be cast into the native 3C framework for analytical tractability. We further illustrate the mobile 3C concepts with practical examples, including a system on broadcast-unicast convergence for massive media content delivery. The mobile 3C design paradigm opens up new possibilities as well as key research problems bearing academic and practice significance.
\end{abstract}

\section{Background and motivation}
Since the introduction of commercial mobile cellular services in 1983 \cite{paper1}, substantial research efforts have been invested in the advancement of mobile networks, most of which, are dedicated to increasing the communication capability of the networks. Indeed, for early generation services such as voice and texting, a higher link capacity or system throughput means better services and more satisfied users, which in turn yields more revenues to the operators. Semantically, the term \textquotedblleft communications" in early generation mobile communications systems (e.g., GSM) can be interpreted as a description of the system's core functionality or its service type - mobile telephony. However, as the mobile network evolves, its service offerings have expanded without boundary. As a matter of fact, it is neither possible nor necessary to list all the services supported by the mobile networks. Consequently, it is only reasonable to interpret the term \textquotedblleft communications" as a depiction of the mobile system's core functionality, i.e., the ability to transmit and receive data over a wireless environment. On the other hand, should the core functionality of future mobile networks be limited to communications? \emph{Are there other capabilities that are equally important and intrinsic to future mobile systems?}

\subsection{The growth of communications (1C) alone is not sustainable}
Throughout the evolution of mobile systems, the emphasis has always be on improving the communication capabilities such as the peak data rate, the minimum delay, and the spectrum efficiency. Early breakthroughs in modulation, channel equalization and multiple-access technologies had fueled the first and second generation cellular networks, whereas advances in channel coding, MIMO, and OFDMA cornstoned the 3G and 4G systems. For the next generation mobile network, the well published technical target is to increase the network capacity by a factor of 1000 beyond 4G, which theoretically could be achieved through the use of \emph{i}) denser basestations, \emph{ii}) additional spectra, and \emph{iii}) improved spectral efficiency. Unfortunately the nature of diminishing return due to the Shannon law deems such gains unsustainable. \textbf{Clearly, alternative solutions are needed in mobile systems to achieve long-term sustainability.}

Unlike the communication resource which is constrained by the available spectrum and power, the computing and caching (i.e., memory) resources are both abundant, economical, and sustainable. The growth of these \textquotedblleft green" resources \cite{paper2} has been following the Moore's low for the past 5 decades with little sign of slowing down. Various attempts have been made to accommodate expanding mobile services through the use of sustainable, non-communication resources. For example, to deliver multimedia contents (which constitute nearly 80\% of the mobile traffic), proactive pushing through data caching has been proposed at both basestations and mobile terminals. In addition to use caching to provide \textquotedblleft individualized" services to mobile users, savings in the communication resources can also achieved through \textquotedblleft computing", in which contents intended for different mobile users are logically processed through coded multicasting or similar techniques \cite{paper3}.

\subsection{Mobile 3C is practically necessary and theoretically canonical}
\emph{Communications-Computing}: The benefits of marrying communications with computing have been studied in various forms within the information-technology communities. In 1984, EI Gamal proposed an open problem \cite{paper4} about the communication complexity of distributed computation in a noisy broadcast network. The minimum of the total number of transmission required to compute any functions in the noisy broadcast network was further derived by Gallager in 1988 \cite{paper5}. In 2000, network coding was proposed by Ahlswede, Cai, Li, and Yeung \cite{paper6}, where algebraic algorithms are applied to data to accumulate the various transmissions in the network. By incorporating simple computing, network coding can improve the effective throughput without extra communication bandwidth. In-network computation was studied in wireless sensor networks by Kumar \cite{paper7} to cooperatively compute a desired function at the destinations.

\emph{Communications-Caching}: Content caching technologies are widely used in content delivery networks (CDN), where contents are pushed and stored in different network locations so that the content closer to the end users can be accessed for QoS optimization. \cite{paper8} investigated the potential of content pushing via broadcasting and showed that exponential gains in the number of users supported can be achieved. Early work on coded multicast \cite{paper3} shows the promise of caching and centralized computing for bandwidth saving. More recently, potential techniques for caching in both the core networks and mobile radio access networks have proposed for 5G cellular systems \cite{paper9,paper10}.
\begin{figure}[t]
\centering
\includegraphics[width=3.5in]{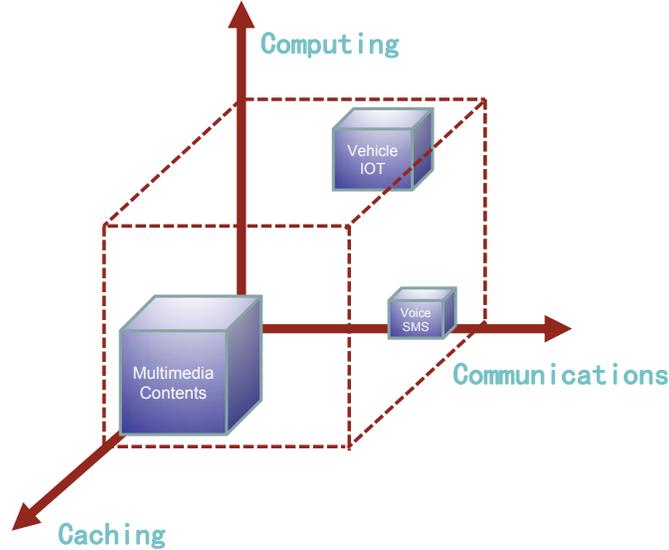}
\caption{The native mobile 3C cube}
\label{figure1}
\end{figure}

To date however, both caching and computing are viewed only as supplementary to the communication functionality of mobile systems, and therefore play a secondary role in the mainstream design efforts. The reasons are probably both historical and academic. Historically, all services offered by legacy mobile systems assuming a \textquotedblleft connection-based" architecture, therefore the quality of services depend entirely on the link capacity of the network. This will no longer be the case in the so-called information-centric networking (ICN) being developed as an alternative Internet architectures \cite{paper11}. In ICN, content caching will become an integral part of network, and the departure from connection-based services allows the wireless network to perform hostless content delivery through in-network caching and named contents, enabling the use of caching and computing throughout the network. On the academic side, the lack of a unified theoretical framework (similar to that of the information theoretical framework for analyzing the communication capacity) handicaps the integration of computing and caching into mobile systems, as their theoretical benefits are far from fully understood. Fundamental questions yet to be answered include but not limited to the following:
\begin{itemize}
  \item   Are caching and computing only supplementary to a mobile system, or should they be treated as intrinsic as communications and thus become the primary resources/functinalities of a mobile network?
  \item   	How can the caching and computing resources be characterized analytically? What are the appropriate metrics to measure the computing and caching resources?
  \item   	How can the mobile network capacity be determined when all 3C resources are included?
\end{itemize}

In this article, we argue in Section II that \textquotedblleft Communications", \textquotedblleft Computing", and \textquotedblleft Caching" (i.e., 3C) make up the three primary resources/functionarlities, akin to the three primary colors, for a mobile system. The three types of resources must be treated as equally essential and intrinsic in order for mobile systems to be scalable and sustainable. Instead of only the communication dimension, the capabilities of future mobile systems should be visualized by a \emph{native mobile 3C cube} as depicted in Fig. \ref{figure1}, in which different services are supported by the 3C core functionalities.

In particular, we attempt to address the above questions by providing a clear definition of each of the 3C resources/functionalities, including their metrics and modes of operations. By formulating \textquotedblleft caching" as a form of non-causal operations and \textquotedblleft computing" as a form of logic operations across information streams, we established the three native and complementary operations for mobile 3C systems.

In Section III, we illustrate the roles of the 3C resources and their impact on mobile system performance with two practical examples. The first one involves a classic cellular system where the 3C resources are exploited at both the basestations and terminals. The second example concerns a space-terrestrial converged network for the purpose of massive content delivery over vast areas. In both cases, the significance of 3C is revealed under the unified view of the mobile system native functionalities, i.e., communications, caching, and computing.

Given that fact that mobile 3C related research is still shaping, we discuss in Section IV key open issues on fundamental theories such as the mobile 3C system capacity limits, as well as practical areas such as the design tradeoffs and protocol supports for the purpose of realizing the full potentials of mobile 3C systems. The paper is then concluded in Section V.
\section{A mobile system in its native form}
The relentless pursuit of higher data rate and system capacity has led to many technological breakthroughs. Owing to space limitations, we are unable to list all the relevant details here; suffice to say that almost all the major advances focused on the \textquotedblleft communications" aspect of the mobile systems. As a matter of fact, in the latest technical outlook for the 5th generation mobile networks, the emphasis on higher communication capacity is highlighted by the so-called \textquotedblleft 5G HyperService Cube" [12], in which the design objectives of three key communication parameters, Links [\emph{Per $\text{Km}^{2}$}], Delay [\emph{ms}], and Throughput [\emph{Kbps/Km}], are explicitly defined \cite{paper12}.

\subsection{The three primary resources in mobile systems}
While the pivotal role of link capability cannot be overemphasized, strictly speaking, only the transmission of real-time data relies \emph{exclusively} on the system's communication functionarlity. Many new services (even "personalized" services such as multimedia content delivery) can be delivered through means other than connection-based transmission. Despite the abundancy of literature on the use of computing and caching in wireless systems, the computing and caching resources have yet to be regarded as essential to a wireless system as communications. On the other hand, the major shift from connection-based services to content-centric services in mobile networks makes it \emph{increasing evident that both caching and computing are essential in providing diverse mobile services, just like the communication functionality.}

Fig. \ref{figure1} presents an alternative view of mobile systems where 3C, i.e., Communications, Caching and Computing, are represented in the \textbf{primary} form for mobile systems. Heuristically the 3C resources form a \textquotedblleft native mobile 3C cube" within which different services can be provided. For example, the traditional real-time, connection-based services such as voice calls rely exclusively on the communication resource, thus operate the X-axis: communications. For multimedia content delivery on the other hand, a service can be provided either through the exclusive use of the communications resource (which is the main driver behind the all for 1000x increase in system capacity), or by leveraging caching (i.e., operating along the Z-axis) and therefore less communications resources \cite{paper13}. Other services such as vehicle networks and Internet-of-Things (IoT) will depend more heavily on the combination of communication and computing resources.
\subsection{The definition of 3C}
In order to analytically characterize the mobile 3C cube, we must first define their metrics and mode of operations so that the impact of 3C can be quantified. In the native 3C form, the capability of the mobile network is determined by three vectors: C\_comm, C\_cache, and C\_comp:
\begin{itemize}
  \item  Communications [C\_comm: \emph{bit/s/Hz}]: the \textquotedblleft communication vector" of the mobile system pertains to its ability to deliver information streams over an imperfect channel with given bandwidth and power. Its capability is measured by the data rate R (unit: bit per second per Hz), whose relation with the system bandwidth and the signal power to noise ratio is well understood by Shannon's capacity formula. \textbf{Note that communication operations, for example, modulations and channel coding, only protect but do not alter the information stream upon delivery.}
  \item  Caching [C\_cach: \emph{bytes}]: the \textquotedblleft caching vector" of the mobile system pertains to its ability to buffer or store certain amount of information (i.e., bit streams) at nodes within the networks. Its capability is typically measured by the memory size (unit: Bytes). While neither protect nor alter the information streams, \textbf{the caching operations introduce non causality or time-reversal into the system}, which in turn, increase the system's capability to deliver information over longer period of time.

  \item Computing [C\_comp: \emph{degree of computing = DoC}]: the \textquotedblleft computing vector" of the mobile system pertains to its ability to perform logic or algebraic operations across information streams. Unlike the communication or caching operations, \textbf{the computing operations do alter the information bits such that additional logic operations must be performed at the destination in order to recover the original information streams.} The computing ability is measured by the number of information streams involved in the operations, which we term the \textquotedblleft degree-of-computing" (DoC).

\end{itemize}

It is important to differentiate the notion of \textquotedblleft caching" and \textquotedblleft computing" as the mobile primary functionalities from those used in the computer science community. Traditionally the computer performance is measured by flops, or floating-point operations per second. However in the mobile 3C system, we are not concern with the amount of computations of a specific algorithm in flops, but rather computing affects the performance limit of the mobile system. In order to establish the theoretical bound on the capacity gains due to the incorporation of computing, we measure the computing functionality by the DoC, or the degree of logical operations. Such definition makes the measure of computing resource algorithm independent. For example, if only one information stream is operated upon, then the DoC = 1 and thus no computing resource is utilized in the system. In this case, the mobile system reduces to the traditional communication mode, i.e., 1C.  On the other hand, when two or more information streams are jointed operated upon, as those performed in network coding, then the DoC = 2+ of computing resource is exploited in the system.

The native 3C resources and their combined effects can be illustrated with the following example on coded multicasting, introduced by Maddha-Ali and Niessen in 2012 \cite{paper3}:

\begin{figure}[H]
\centering
\includegraphics[width=3in]{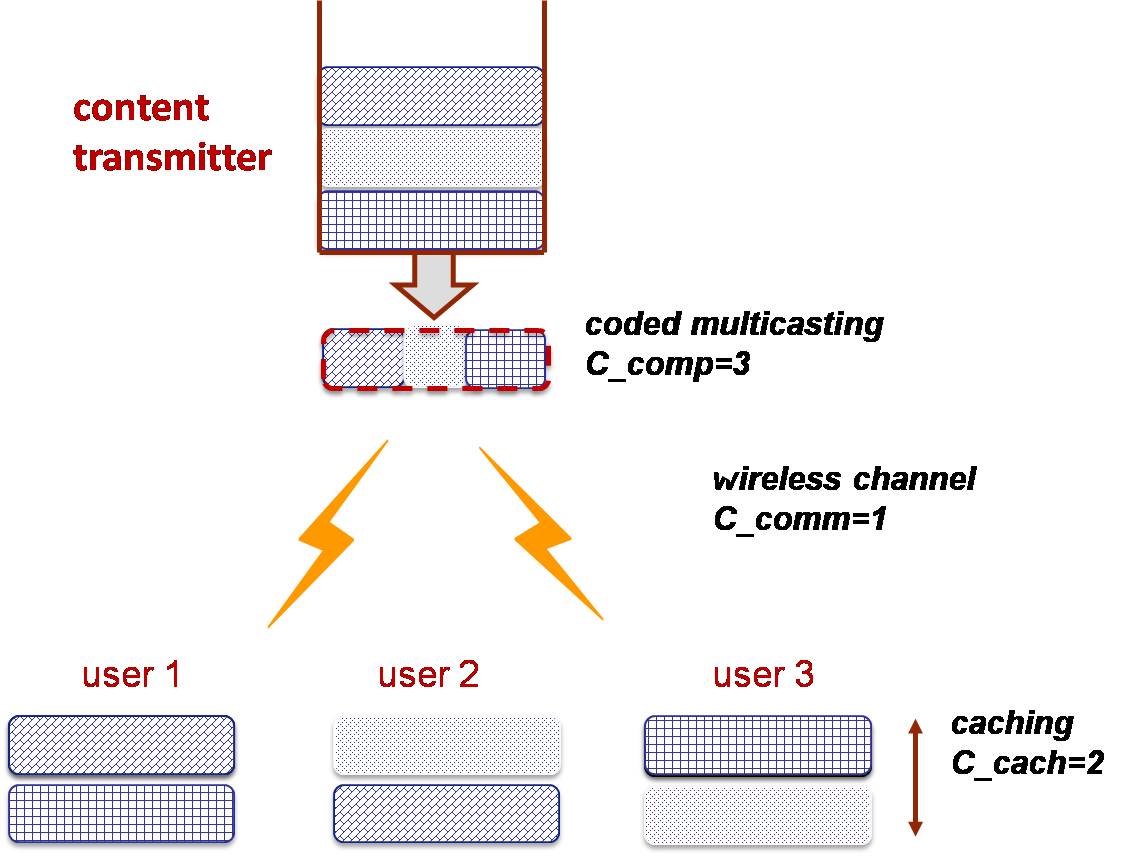}
\caption{3C in wireless coded multicasting}
\label{figure2}
\end{figure}

Fig. \ref{figure2} depicts a scenario in which contents are delivered to 3 mobile users over a shared wireless channel. Given a fixed amount of power and bandwidth, the 1st primary resource of the wireless system, i.e., communications, in normalized to unity: C\_comm =1. For a conventional wireless system with no additional resources in the system, the instantaneous sum rate of three mobile users, which is also the rate at which the maximum amount of contents can be delivered, is upper limited at 1:
\begin{equation}
R_{1}+R_{2}+R_{3}\leq1.
\end{equation}%

Now consider the incorporation of the 2rd primary resource: caching at the user ends. In this particular example, the caching size of all users is set at 2, i.e., C\_cach=2. As mentioned above, caching increases the system capacity by introducing non-causality in the system. For example, the amount of communication resources required to deliver 1 bit of content over $T$ seconds is $r=1/T$ bit/s/Hz. With caching on the other hand, the content delivery process can start at $-T$ seconds or earlier. Consequently the required communication resources to deliver the same content reduces to $r=1/|[-T~~T]| = 1/2T$, effectively doubles the content delivery capability without additional communication resources.

Now consider the addition of the 3rd primary resource: computing. In the case of code multicast, computing is invoked to perform logical operations across content bits intended for all three mobile users, which means C\_comp=3. This \textquotedblleft computing gain" (or the centralized coding gain as defined in \cite{paper3}) is calculated as

\begin{equation}
\frac{1}{1+N/(C\_comp\times C\_cach)},
\end{equation}%
where $N$ denotes the number of contents in a server. Interestingly, the computing gain is nearly linear with both C\_comp and C\_cach, as opposite to logarithmic in the case of communication resources. When $N\gg C\_comp\times C\_cach$, the computing gain becomes $C\_comp\times C\_cach/N$, which means the system enjoys linear gain with respective to both C\_comp and C\_cach.

Three important observations can be drawn from the above example:
\begin{enumerate}
  \item While each of the 3 primary resources plays a different role in the mobile systems, it is their combined effects that lead to the gains in system capacity. The optimum operating point depends on the particular service offering and the 3C resource constraints.

  \item In contrast to the communication resource (bit/s/Hz), which value is a real number, both caching (Bytes) and computing (DoC) are measured in integers. Such definitions not only decouple caching and computing from bandwidth and power, but also make the theoretical analysis more tractable.

  \item Observe that in this particular case, the capability of content delivery is proportional to C\_cach and C\_comp. In other cases, caching has been shown to provide exponential gains in terms of system's capacity to deliver massive contents.
\end{enumerate}

\section{From theory to practice}
In reality, many attempts on 3C collaborations already have been made in the mobile systems. In this section we cast two practical mobile systems into the 3C framework. The reformulation is not intended to reinvent the existing solutions, but to offer an alternative perspective and more insights under the 3C model.

\subsection{3C in mobile cellular applications}

\begin{figure}[t]
\centering
\includegraphics[width=5.5in]{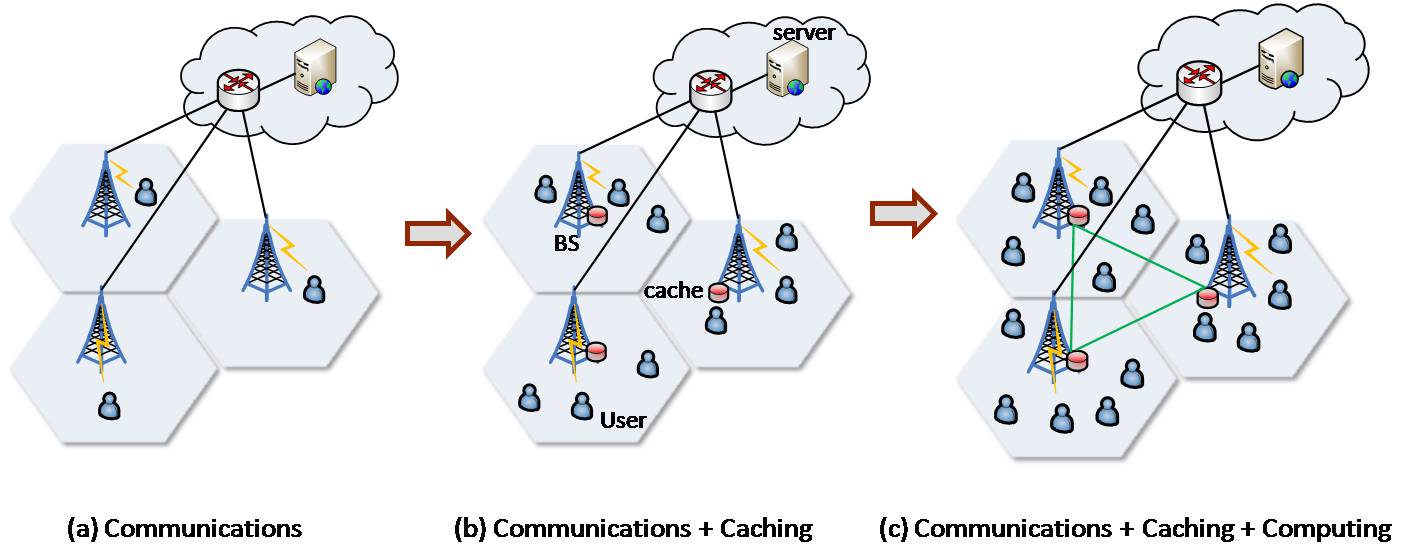}
\caption{3C in mobile cellular systems}
\label{figure3}
\end{figure}

Fig. \ref{figure3} shows a typical mobile cellular system in which the 3C resources are exploited in different stages. A simple mobile 1C network, as illustrated in Fig. \ref{figure3}(a), only utilizes the communication resource and provides no caching ability at either the base stations or at the user terminals. Functioning solely as a transceiver, each base station has to serve users' requests instantaneously over wireless channels and relay them directly to the core networks, resulting in potential bottlenecks at both the backhaul and over-the-air traffics.

Fig. \ref{figure3}(b) depicts the scenario in which the caching resource is added at the basestations. Accordingly, the system expands from 1C to 2C, which allows the network to store the popular contents at the basestations. Now users requests can be directly served via the content cached in the \emph{local} storage instead of the backhaul. This caching mechanism dramatically decreases the duplicated transmissions between the base station and the core network.

Basic on the definition of DoC in Section II-B, the DoC of both Fig. \ref{figure3}(a) and Fig. \ref{figure3}(b) is only 1. That is, no computing resources is exploited in these settings. The multicell network can further expand its capacity by increasing the DoC in a way illustrated in Figure 3 (c) where the basestations are inter-connected with a backhaul link. Specifically, the basestations collaborate and share the caching information among neighboring basestations. As a result, each base station can serve a content from its local storage, or from a neighboring basestation, or from the core network. The DoC in this case is defined as the number of collaborative basestations from which cached contents can be shared.

\begin{figure}[t]
\centering
\includegraphics[width=4.5in]{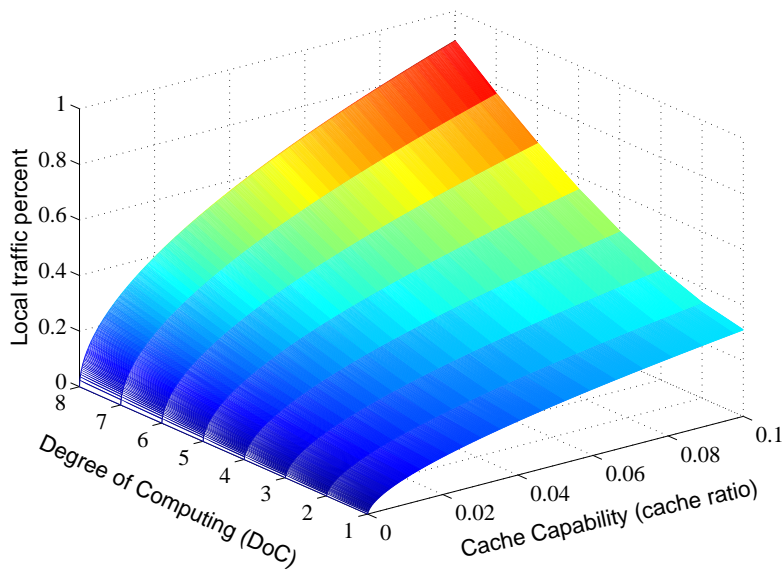}
\caption{Local traffic percent vs. DoC and Caching Capability}
\label{figure4}
\end{figure}

The incorporation of 3C has positive impact on all system performance parameters, including the total number of terminals served under certain QoS constraints, the average delays for content delivery, and bandwidth savings etc. Fig. \ref{figure4} illustrates one of the performance improvements by measuring the local traffic percent with respect to the DoC and the cache ratio. The local traffic percent is defined as the ratio of the amount of contents served from local storage to the total amount of contents served. As one can see the local traffic percent grows with increasing caching capability and the DoC. The higher the DOC is, the larger benefit caching brings. On the other hand when the network has small caching capacity, e.g., caching ratio $<0.01$, the DoC increase only leads to insignificant performance gains. The results reconfirm the importance of simultaneous exploitation of all 3Cs in mobile networks.

\subsection{3C in converged mobile networks}
In the subsection, we describe a novel network being designed under the mobile 3C framework. The application of interest here is of massive multimedia content delivery for vast-area mobile terminals, such as vehicles, drones, trains, ships and sensors, etc. The unique application requires the soultion to be ubiquitous, scalable, and at the same time, economical. The existing infrastructures that are readily available include the cellular networks, the terrestrial broadcasting, the satellite broadcasting, and the satellite communications systems. Unfortunately none of these systems, when works independently, can simultaneously meet all the stringent requirements. The cellular networks for example, while capable of providing extremely high data rates, only covers about 20\% of the landmass, and thus are inadequate to serve vast areas. On the other hand, the satellite-based solutions offer much broader coverage but are either limited in capacity or intractability.

The above challenges are being tackled by a so-called \textquotedblleft space-terrestrial, unicast-broadcast convergence" (STUBC) system under development, see Figure 5. The STUBC maximizes the strengths of individual mobile networks and overcome their limitations. Without getting into details of the converged networks, it is suffice to describe the solution as "over-the top content provision" (OCP) through dynamic utilization of the 3C resource within the heterogeneous satellite and terrestrial networks. By exploiting the caching and computing resources, the strategy can potentially offload 80\% the multimedia traffics onto the broadcast networks and provides the rest of services though unicasting \cite{paper8}.

\begin{figure}[t]
\centering
\includegraphics[width=4.5in]{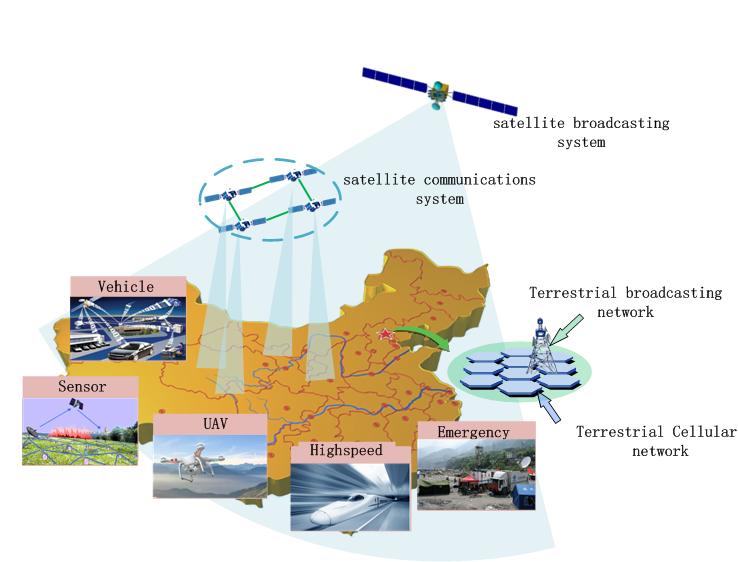}
\caption{3C-enabled Convergence Networks}
\label{figure5}
\end{figure}

In order to collaboratively deliver contents across different networks, the STUBC system augments the existing communication protocols with two modules for the purpose of exploiting the heterogeneous 3C resources:
\begin{itemize}
  \item  Task Representation Module (TRM): The essence of the converged mobile 3C network is to distribute content objects according to OCP's requirements by leveraging all 3C resources within the heterogeneous network. The purpose of the TRM is to provide task profiles specifying the preferences of each content object distribution subtask, including the types and the amount of resources needed, QoS constraints and preferred management interfaces. The OCP could understand the popularity of each content object, among other characteristics, which could in turn help the OCP generate the task profile.

  \item  Service Negotiation Module (SNM): The OCP attempts to deliver its modularized content objects to users by issuing tasks, with the users' context and popularity of each object taken into account, to the heterogeneous network. Since the 3C resources are controlled by individual networks, a SNM resides in between the OCP and the wireless heterogeneous networks is needed to negotiate the use of 3C resources within each network for distributing contents, based on the \emph{task profile and the service level agreement} provided by individual wireless networks. After the negotiation process, the modularized objects can then be distributed over the converged mobile network.
\end{itemize}

\section{Open Issues}
The mobile 3C design paradigm opens up new possibilities as well as key research problems bearing academic and practice significance. On the academic side, the classic information theoretical model which has anchored many fundamental breakthroughs in mobile technologies cannot be easily extended to mobile 3C systems. Similarly the wireless communications protocols established for early generations of mobile networks will be insufficient in exploiting the 3C resources.

\begin{description}
  \item[A.] \textbf{The \textquotedblleft 3C capacity"}: One of the most challenging tasks associated with mobile 3C research is to determine its theoretical capacity. The problem is rather involved even for relative simple scenario such as the one depicted in Fig. \ref{figure3}(b). The classic information theoretical model that addresses \emph{instantaneous} rate regions is not directly applicable here, as it cannot adequately address the caching-induced non-causality within the system. As a matter of fact, our recent study shows that the conventional notion of rate capacity is unsuitable to describe the network's capability to deliver non-private contents for multiple users \cite{paper13}.

\par
      To measure the impact of caching in content delivery, a so-called \textquotedblleft content rate" was introduced to better measure the rate at which the amount of contents is successfully delivered to users over the shared wireless channel \cite{paper13}.

      \textbf{Definition 1}: Given a fixed amount of wireless resource $B\times T$ (bandwidth and time), the content rate $R_{C}$ is defined as the total amount of service packages successfully delivered to the users: $R_{C}=\frac{\sum_{k=1}^{K}|y_{k}(t_{k})|}{B\times T}$.

      \par

      The study reveals that when the contents objects are neither private nor common, the maximum content rate is achieved when the caching size reaches infinity, whereas the minimum content rate is determined by the classic information rate bound with zero caching.

      \par

      There exists a large literature on how the incorporation of computing affect the capacity of the mobile network. Within these literature, however, there exists a contrast over the explicit role of computing in capacity calculation. One subset of approaches is that of network coding where the algebraic operations (i.e., computing) are distinguished from the communication operations \cite{paper6}. Others approaches, such as those of BTS collaborative transmission \cite{paper14} and distributed MIMO \cite{paper15}, do not decouple computing vs. communications, as the two types of operations (i.e., channel coding and logic operations across information streams) are intertwined. Clearly, a unified capacity analysis in which the 3C resources are represented in the canonical form (as in Fig.\ref{figure1}) will carry tremendous theoretical value.

  \item[B.] \textbf{The 3C tradeoffs}: In early sessions we show through examples that the same types of services can be accomplished through different combinations of 3C resources, represented by an operating subspace inside the 3C cube. For example, volume contents can be unicasted to mobile users upon requests as is performed in connection-based wireless networks, leading to the maximum usage of the communication resources but zero caching or computing. Alternatively, selected contents (especially the popular ones) can be proactively pushed and cached at basestations and/or the user terminals, effectively trading the scarce communications resources with the more abundant and sustainable caching and computing resources.

      Unfortunately for each service type (see Fig. \ref{figure1}), there exists such tradeoffs which has to be individual analyzed against its associated performance measures and resource constraints. While it is nontrivial to determine the \textquotedblleft optimum" tradeoff in each cases, one could argue that the Moore's law strongly favorites the use of caching and computing resources over the communication resource, at least in the foreseeable future.

  \item[C.] \textbf{Software defined mobile 3C networks}: By treating communications, computing and caching all as primary and intrinsic resources within mobile systems, a fundamentally different approach need to be adopted in the design of future generation networks. Indeed, the attempt to dynamic allocate the communication resources (e.g., software defined networks - SDN) should be extended to both the caching and computing resources, entailing an entirely different set of signalings and protocols. While such an exercise is academically feasible, its practical implementation is particularly challenging for two reasons.

      First of all, none of the existing communication protocols allows direct access to the contents themselves, and therefore hinders the exploitation of distributed caching within the network. The inability of the mobile systems to sufficiently exploit the content characteristics is in part attributable to the fact that traditional connection-based protocols. The need of fully utilizing the caching resources for content distribution has motivated the development of future network architectures based on the named data objects (NDOs), which is commonly known as the information-centric networking (ICN) \cite{paper11}. However, the full integration of ICN with mobile network protocols seems to be in distant future.

      The second challenge ahead is mostly algorithmic. Conceptually, a software defined mobile networks with rich 3C resources would lead to substantial gains. However the dynamic allocations of 3C resources, most of them distributed, has significant implications on the signaling overhead, algorithm implement ability and network stability. Both fruitful research activities and significant industrial advancements along this direction can be anticipated in coming years.
\end{description}

\section{Conclusion}
In this article, we have presented an alternative framework for mobile systems in which communications, caching and computing (3C) are identified as the three primary resources or functionalities. By formulating \textquotedblleft caching" as a form of non-causal operations and \textquotedblleft computing" as a form of logic operations across information streams, we established the three native and complementary operations for mobile 3C systems. The fact that two of the primary resources are scalable and sustainable (i.e., not subject to the bandwidth and power constraints) suggests a fundamentally different design paradigm for future mobile systems. We have shown with several examples that, the unified view of the mobile system functionalities provides deep insights on how the 3C resources could be utilized to impacts the overall system performance. This led to a discussion of open issues for 3C mobile system, not only in fundamental theories such as the mobile 3C system capacity, but also in practical areas such as design tradeoffs and protocol supports for realizing the full potentials of mobile 3C systems.

\bibliographystyle{IEEEtran}
\bibliography{content_journal}

\end{document}